\documentclass{emulateapj}
\usepackage{amsmath}
\usepackage{mathptmx}

\newcommand{\unitspace}{\ensuremath{\,}}
\newcommand{\usp}{\unitspace}
\newcommand{\numberspace}{\ensuremath{\;}}
\newcommand{\nsp}{\numberspace}
\newcommand{\unitstyle}[1]{\ensuremath{\mathrm{#1}}}
\newcommand{\power}[2]{\ensuremath{{#1}^{#2}}}

\newcommand{\kilo}{\unitstyle{k}}
\newcommand{\Mega}{\unitstyle{M}}

\newcommand{\cm}{\unitstyle{cm}}
\newcommand{\gram}{\unitstyle{g}}

\newcommand{\meter}{\unitstyle{m}}

\newcommand{\second}{\unitstyle{s}}

\newcommand{\Kelvin}{\unitstyle{K}}

\newcommand{\grampercc}{\gram\usp\power{\cm}{-3}} 
\newcommand{\grampersquarecm}{\gram\usp\power{\cm}{-2}} 

\newcommand{\GramPerSc}{\grampersquarecm}
\newcommand{\erg}{\unitstyle{ergs}} 
\newcommand{\ergs}{\erg}
\newcommand{\ergspersecond}{\erg\unitspace\power{\second}{-1}}

\newcommand{\fermi}{\unitstyle{fm}} 
\newcommand{\eV}{\unitstyle{eV}}        
\newcommand{\MeV}{\Mega\eV} 

\newcommand{\Msun}{\ensuremath{M_\odot}}

\newcommand{\yr}{\unitstyle{yr}}        

\newcommand{\NA}{\ensuremath{N_\mathrm{\!A}}} 
\newcommand{\mb}{\ensuremath{m_\mathrm{u}}} 

\newcommand{\satellite}[1]{\emph{#1}}

\newcommand{\beppo}{\satellite{BeppoSAX}}

\newcommand{\rxte}{\satellite{RXTE}}

\newcommand{\ee}[1]{\ensuremath{\times 10^{#1}}}

\newcommand{\source}[3]{#1~#2$#3$} 
\newcommand{\nuclei}[2]{\ensuremath{\mathrm{^{#1}#2}}}
\newcommand{\CC}{\ensuremath{\nuclei{12}{C}+\nuclei{12}{C}}}
\newcommand{\Mdot}{\ensuremath{\dot{M}}}
\newcommand{\MdotEdd}{\ensuremath{\Mdot_{\mathrm{Edd}}}}
\newcommand{\rhoc}{\ensuremath{\rho_\circ}}
\newcommand{\rhob}{\ensuremath{\rho_\mathrm{b}}}
\newcommand{\rhonuc}{\ensuremath{\rho_\mathrm{nuc}}}
\newcommand{\Tb}{\ensuremath{T_\mathrm{b}}}

\newcommand{\EN}{\ensuremath{\varepsilon_{\mathrm{nuc}}}}
\newcommand{\Enu}{\ensuremath{\varepsilon_\nu}}
\newcommand{\ECC}{\ensuremath{\varepsilon_{12,12}}}
\newcommand{\Ecool}{\ensuremath{\varepsilon_\mathrm{cool}}}
\newcommand{\Lnuc}{\ensuremath{L_\mathrm{nuc}}}
\newcommand{\grampersecond}{\ensuremath{\gram\usp\power{\second}{-1}}}

\newcommand{\meanA}{\ensuremath{\langle A\rangle}}
\newcommand{\meanZ}{\ensuremath{\langle Z\rangle}}
\newcommand{\meanZZ}{\ensuremath{\langle Z^2\rangle}}

\begin{document}
\title{Superburst Ignition and Implications for Neutron Star Interiors}
\author{Edward F. Brown}
\affil{Department of Physics and Astronomy and the Joint Institute for
Nuclear Astrophysics, Michigan State University, East Lansing, MI 48824}
\email{ebrown@pa.msu.edu}
\journalinfo{ApJ Letters, in press}

\begin{abstract}
  Superbursts are thought to be powered by the unstable ignition of a
  carbon-enriched layer formed from the burning of accreted hydrogen and
  helium.  As shown by Cumming \& Bildsten, the short recurrence time
  hinges on the crust being sufficiently hot at densities $\rho >
  10^9\usp\grampercc$.  In this Letter, we self-consistently solve for
  the flux coming from the deep crust and core.  The temperature where
  the carbon unstably ignites is only weakly sensitive to the
  composition of the ashes of H/He burning, but does depend on the
  thermal conductivity of the inner crust and the neutrino emissivity of
  the core.  The observed superburst recurrence times and energetics
  suggest that the crust thermal conductivity is low, as if the crust
  were amorphous instead of crystalline.  If the conductivity is higher,
  such as from a lattice with impurities, then matching the superburst
  properties require that the neutrino emissivity be not stronger than
  modified Urca.  Observations of superbursts---energetics, recurrence
  times, and cooling times---therefore complement observations of
  isolated cooling neutron stars and soft X-ray transients in
  constraining properties of dense matter.  Perhaps the most interesting
  object in this regard is \source{KS}{1731}{-260}, which produced a
  superburst during its protracted accretion outburst but had a rapidly
  declining quiescent luminosity.
\end{abstract}


\section{Introduction}\label{sec:introduction}

With the launch of \beppo\ and \rxte, long-term monitoring of low-mass
X-ray binaries became feasible.  One fruit of this was the discovery of
superbursts: these have a rapid rise and long decay, analogous to type I
X-ray bursts, but are roughly a thousand times more energetic
\citep{cornelisse.ea:longest,.kuulkers.ea:bepposax,kuulkers:superoutburst,kuulkers.ea:ks1731-superburst,kuulkers:gx31_super,strohmayer.brown:remarkable,wijnands:recurrent}.
The recurrence time is of order a year; for one source,
\source{4U}{1636}{-53}, two superbursts were observed separated by
4.7\nsp\yr\ \citep{wijnands:recurrent}.  Although most superbursts are
observed at mass accretion rates 0.1--0.3\nsp\MdotEdd, $\MdotEdd \approx
10^{18}\usp\grampersecond$ being the Eddington accretion rate, recently
\citet*{.cornelisse.ea:superbursts} detected superbursts from the
rapidly accreting source \source{GX}{17}{+2}.

Despite their similarities to type I X-ray bursts, the fuel for the
superbursts was not immediately apparent.  \citet{taam78:_nuclear} had
originally investigated the thermal stability of a pure \nuclei{12}{C}
layer and found that such a layer would unstably ignite when the
accumulated mass was $\sim 10^{27}\nsp\gram$.  This produces a very
energetic burst, $E\sim 10^{44}\usp\ergs$, with a recurrence time of
$300\nsp\yr\nsp(\dot{M}/0.1 \usp\MdotEdd)$.  \citet{brown98}
demonstrated that carbon flashes with greatly reduced recurrence times
could occur on a neutron star (NS) accreting pure He at a \emph{locally}
super-Eddington rate, such as might happen on an accreting X-ray pulsar.
This explanation did not fit the observed superbursts because none of
the sources showed pulsations in the persistent emission and, with the
exception of \source{4U}{1820}{-30}, all were thought to accrete a H/He
mixture that burned to matter with $A\sim 100$
\citep{schatz99,schatz.aprahamian.ea:endpoint}.

\citet{cumming.bildsten:carbon} suggested a resolution to this puzzle.
They noted that only a small \nuclei{12}{C} mass fraction
$X(\nuclei{12}{C}) \sim 0.1$ is sufficient to produce a thermonuclear
runaway, and that the reduced conductivity of a heavy element mixture
(e.g., 0.9:0.1 \nuclei{104}{Ru}:\nuclei{12}{C}) would produce a steeper
thermal gradient for a fixed flux emergent from the deeper crust.  The
flux from the deep crust is powered by electron captures, neutron
emissions and pycnonuclear reactions
\citep{haensel90a,haensel.zdunik:nuclear} that release roughly
1.4\nsp\MeV\ per accreted nucleon; the outward-directed luminosity from
the deep core is roughly $\Lnuc \lesssim 0.1\usp\MeV (\Mdot/\mb)$, for
steadily accreting NSs with modified Urca neutrino cooling in their
cores \citep{brown:nuclear}.  Here \mb\ is the mean nucleon mass.  Under
these conditions, superbursts would ignite for an accumulated mass $\sim
10^{25}\nsp\gram$, which is consistent with the observed energetics and
recurrence times.  The superbursts observed from \source{GX}{17}{+2} are
consistent with this scenario \citep{.cornelisse.ea:superbursts}.

In this Letter, we calculate the thermal structure of the entire crust
using a range of core neutrino emissivities and crust thermal
conductivities.  The flux in the carbon-enriched layer is not fixed, but
increases with $\meanA/\meanZZ$, where \meanA\ and \meanZZ\ are the mean
nuclear mass and squared nuclear charge of the carbon-enriched layer.
As a result, the temperature at which superbursts ignite is relatively
insensitive to the dominant isotope formed during H/He burning.
Furthermore, the temperature of this layer is sensitive to the nature of
the thermal conductivity and neutrino emissivity of the deep crust and
core.  As a result, the superburst properties---recurrence times,
energies, and cooling times---open a new view into the interior physics
of NSs, complementary to observations of isolated cooling neutron stars
\citep[for a recent review, see][]{yakovlev.pethick:neutron} and soft
X-ray transients
\citep{rutledge.ea.01:ks1731,yakovlev.levenfish.ea:thermal,yakovlev.levenfish.ea:coldest}.
In particular, the observed energetics and recurrence times agree with
the scenario of \citet{cumming.bildsten:carbon} only if the nuclei in
the NS crust are disordered or amorphous.

\section{The Thermal Structure of the Crust}\label{sec:thermal-structure}

To calculate the crust thermal structure, we use the method outlined in
\citet{brown:nuclear} with a few modifications.  First, we take the core
neutrino emissivity as a free parameter, rather than compute it
self-consistently for a given equation of state (EOS).  This is because
our goal is to study the crust thermal structure for a given
\emph{average} core neutrino emissivity.  To simplify our calculations,
we follow the approach of \citet{yakovlev.haensel:what} and use an
analytical density prescription $\rho = \rhoc[1-(r/R)^2]$
\citep{tolman:1939}, where $\rhoc = 15 (8\pi)^{-1} M R^{-3}$ with $R =
(2GM/c^2) \left[1 - \left( 1+z \right)^{-2} \right]^{-1}$ and $1+z =
[1-2GM/(Rc^2)]^{-1/2}$.  This prescription gives a reasonable
approximation to the $\rho(r)$ found using modern nuclear EOS's
\citep[see][and references therein]{lattimer.prakash:neutron}.  With
this choice for the EOS, the NS's mass $M$ and redshift $z$ specify its
mechanical structure.  We locate the crust-core boundary by a Maxwell
construction using an EOS for the inner crust \citep{negele73} and the
core \citep*{akmal98}, which gives the transition density $\rhob =
1.6\ee{14}\usp\grampercc$.

With the core mass and radius specified, we compute the structure of the
crust by integrating the general relativistic equation of hydrostatic
balance from $P(\rho=\rhob)$ to $P = g\times 10^9\nsp\GramPerSc$.  We
then solve the thermal equations
\begin{eqnarray}
  \label{eq:T}
  e^{-\nu/2}\partial_r\left(e^{\nu/2} T\right) &=& 
  -(1+z) \frac{L}{4\pi r^2\, K} \\
  \label{eq:L}
  e^{-\nu}\partial_r\left(e^\nu L\right) &=& 
  4\pi r^2 \left(1+z\right) \left(\EN - \Enu\right)
\end{eqnarray}
by a relaxation method \citep{pre92}.  Here $T$ and $L$ denote proper
values and $e^\nu \approx (1+z)^{-2}$.  The condition at the crust/core
boundary is $L(\rho = \rho_b) + L_\nu = 0 $: the heat conducted into the
core must be re-radiated as neutrinos.  At a column depth $y =
10^9\nsp\GramPerSc$ we set $T = 2.5\ee{8}\nsp\Kelvin$, in accordance
with estimates for mixed H/He ignition
\citep{cumming.bildsten:rotational}.

The crust thermal profile is controlled by the EOS, the volumetric
neutrino emissivity \Enu\ and nuclear heating rate \EN, and the thermal
conductivity $K$.  The crust EOS has contributions from degenerate,
relativistic electrons, for which we use a tabulation of the Helmholtz
free energy\footnote{Extension to nuclear densities courtesy F. Timmes}
\citep{timmes.swesty:accuracy}; from ions, for which we use the
free-energy fits of \citet{farouki93}; and from free neutrons, for which
we use a compressible liquid-drop model \citep{mackie77}.  The neutrino
emission in the inner crust is dominated by neutrino pair bremsstrahlung
\citep{maxwell79:_neutr}, for which we use the rate calculated by
\citet*{haensel96}.  As in \citet{brown:nuclear}, we do not resolve the
reaction layers in the deep crust, but instead use a distributed \EN\ 
normalized so that the total luminosity from the deep crust reactions is
$\Lnuc = (1.4\usp\MeV)\times(\dot{M}/\mb)$.

Heat is conducted by electrons, and the thermal conductivity is given by
the Wiedemann-Franz law, with the conductivity proportional to an
electron-ion relaxation time $\tau$, for which we use the fitting
formula of \citet{potekhin99:_trans}, in the mean-ion approximation.  A
substantial uncertainty in computing the transport coefficients of the
crust is its composition.  During the lifetime of a low-mass binary, it
is possible to replace a substantial portion of the $\sim 0.01\nsp\Msun$
crust.  The accreted hydrogen and helium burn via the rp-process
\citep{wallace81:_explos}; both steady-state \citep{schatz99} and
time-dependent, one-zone \citep{schatz.aprahamian.ea:endpoint}
calculations predict that the ashes are a broad distribution of ions
peaking with $\meanA\approx 100$.  Recent calculations of
one-dimensional models of episodic accretion and unstable H/He ignition
for a range of accretion rates and metallicities find that repeated
bursts drive the mean weight of the ashes to $\meanA\approx 64$
\citep{woosley.heger.ea:models,fisker.brown.ea:extracting}.  For
definiteness, we use the compositional profile of
\citet{haensel.zdunik:nuclear}, which presumes that the ashes of H/He
burning have $\meanA \approx 100$; in the inner crust the composition is
similar, however, to that of \citet{haensel90a}, which assumed the ashes
of H/He burning are \nuclei{56}{Fe}.

\begin{figure}[htb]
  \includegraphics[width=\hsize]{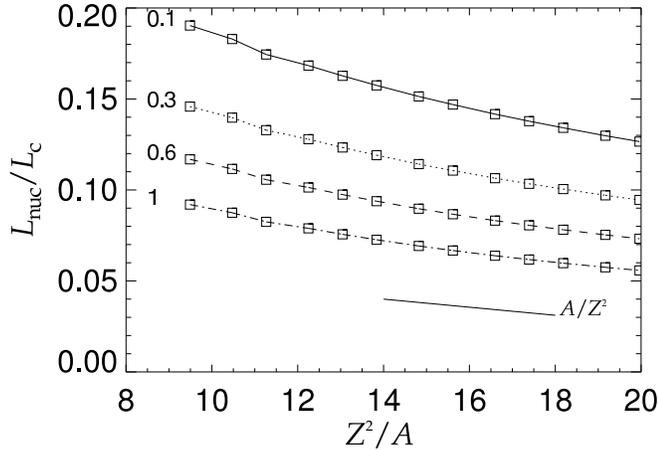}
  \caption{The emergent luminosity from the deep crust,
    scaled to $L_\mathrm{C} = 1\nsp\MeV\times(\dot{M}/\mb)$ for
    $\dot{M}_{17} = 0.1$, 0.3, 0.6, and 10.0 (see text).  This is for a
    crust with low thermal conductivity.}
\label{fig:Z2Ascale}
\end{figure}

As pointed out by \citet{cumming.bildsten:carbon}, the composition of
the carbon-enriched layer can have a strong effect on the temperature
profile.  In this layer the ions are not crystallized and the
conductivity scales roughly as $\meanA/\meanZZ$.  As a result, one would
expect a steeper gradient for a heavier ion mixture and a given flux.
The flux is not fixed, however, and must be determined
self-consistently.  Figure~\ref{fig:Z2Ascale} shows the luminosity
emergent from the crust for different values of $Z^2/A$ in the layer
$10^9\nsp\grampercc < y < 10^{13}\nsp\grampercc$.  The connected squares
mark the results of the calculations; the solid line indicates the slope
$A/Z^2$.  The luminosity is scaled by a fiducial value, $L_\mathrm{C} =
1\nsp\MeV\times (\dot{M}/\mb) =
9.6\ee{34}\usp\ergspersecond\nsp\dot{M}_{17}$, and the numbers beside
each line indicate the accretion rate $\dot{M}_{17} =
\dot{M}/(10^{17}\usp\grampersecond)$.

To understand why the thermal gradient depends weakly on the composition
of the layer below the H/He burning shell, consider a simple two-zone
model of the crust, with the heating from the pycnonuclear reactions at
the interface between two zones.  When most of the heat generated in the
crust flows into the core, the maximum temperature in the crust, and
therefore the thermal gradient between the H/He burning layer and the
electron capture layers, is roughly fixed.  This approximate constancy
in the temperature gradient implies that the combination $L
\meanZZ/\meanA$ is roughly constant (Fig.~\ref{fig:Z2Ascale}).
\emph{The superburst ignition temperature depends weakly on the
  composition of the carbon-enriched layer.}  At lower accretion rates
and smaller $Z^2/A$, this simple argument weakens, as the luminosity in
the outer layers becomes a larger fraction of \Lnuc.  For use in future
computations, we fit the luminosity in the carbon-enriched layer as a
function of $\dot{M}_{17}$ and $Z^2/A$ for $1 < \dot{M}_{17} < 10$ and
$8 < Z^2/A < 20$.  We write $L = q\dot{M}\NA (1\usp\MeV) =
(9.65\ee{34}\usp\ergspersecond)\times q\dot{M}_{17}$, with
\begin{equation}
  \label{eq:fit}
  q = 0.386 \left(\frac{Z^2/A}{12}\right)^{-0.6}
  \left(1 - 0.562\dot{M}_{17}^{0.15}\right).
\end{equation}
This fit has an rms error of 1.7\%, with a maximum error of 5.1\% for
$\dot{M}_{17} = 10$ and $Z^2/A = 20$.

\section{Sensitivity to the physics of the deep crust and core}
\label{sec:results}

To explore the effects of the core and crust microphysics on the carbon
ignition, we choose a core model with $M= 1.6\usp\Msun$ and $z = 0.35$,
which implies that $R = 10.5\nsp\kilo\meter$ and $\rhoc = 6.2\rhonuc =
1.7\ee{15}\nsp\grampercc$.  Here $\rhonuc/\mb = 0.16\nsp\fermi^{-3}$ is
the fiducial nuclear density.  Again following
\citet{yakovlev.haensel:what}, we use two cases of the core neutrino
emissivity.  The first case is $\rho\Enu =
3.0\ee{19}\nsp\ergspersecond\usp\cm^{-3} (T/10^9\nsp\Kelvin)^8$
throughout the core.  This corresponds to a modified Urca process
somewhat suppressed by superfluidity.  For the second case, we add an
enhanced emissivity $\rho\Enu = 10^{26}\nsp\ergspersecond\usp\cm^{-3}
(T/10^9\nsp\Kelvin)^6$ for $\rho > 5.0\rhonuc$.  This models the
operation of an enhanced neutrino emissivity from, e.g., a direct Urca
process above a density threshold \citep{lattimer91}.  These choices
change the core temperature by an order of magnitude for a given
accretion rate: the proper temperature at the base of the crust is $\Tb
\approx 4.6\ee{8}\nsp\Kelvin (\Mdot/\MdotEdd)^{1/8}$ for the first case
and $\Tb \approx 3.9\ee{7}\nsp\Kelvin (\Mdot/\MdotEdd)^{1/6}$ for the
second.  The core temperature is insensitive to the exact value of the
density threshold.

In addition to the neutrino emissivity of the core, we also choose two
cases for the crust thermal conductivity.  Calculations of the ionic
state of the crust usually presume a single-species lattice, with
possible impurities.  The actual crust is formed, however, from the
processing of rp-process ashes, and it is not known if the ions form an
ordered lattice.  We therefore pick two cases for comparison.  First,
that the lattice is disordered; we estimate the electron-ion relaxation
time $\tau$ by setting the structure factor to unity
\citep[see][]{itoh93}.  In this case $\tau^{-1}$ is independent of $T$
and proportional to $Z^2/A$.  For the second case, we compute $\tau^{-1}
= \tau^{-1}_\mathrm{e,ph} + \tau^{-1}_\mathrm{e,imp}$, where
$\tau^{-1}_\mathrm{e,ph}$ is the frequency of electron-phonon scattering
and $\tau^{-1}_\mathrm{e,imp}$ is the frequency of electron-impurity
scattering \citep{potekhin99:_trans}, for which we set $Q = \langle Z^2
-\meanZ^2\rangle = 100$.  This is typical of the rp-process ashes, but
it is larger than that of matter burned in a superburst
\citep*{schatz.bildsten.ea:photodisintegration-triggered} or formed at
the birth of the neutron star \citep{jones:heterogeneity}.

\begin{figure}[htb]
  \includegraphics[width=\hsize]{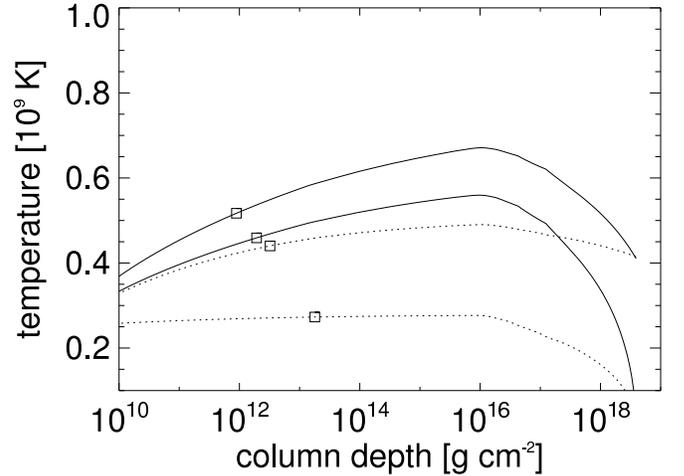}
  \caption{Crust temperatures for $\Mdot =
    3.0\ee{17}\nsp\grampersecond$.  The squares indicate where a mixture
    with $X(\nuclei{12}{C}) = 0.1$ would unstably ignite.  The solid
    lines are for the crust thermal conductivity dominated by
    scattering; the dotted lines indicate a conductivity that is a mix
    of phonon and impurity scattering.  The solutions with higher core
    temperatures correspond include only modified Urca processes; the
    solutions with a cold core temperature include in addition an
    enhanced neutrino emission.}
\label{fig:mdot0.3}
\end{figure}

\begin{figure}[htb]
  \includegraphics[width=\hsize]{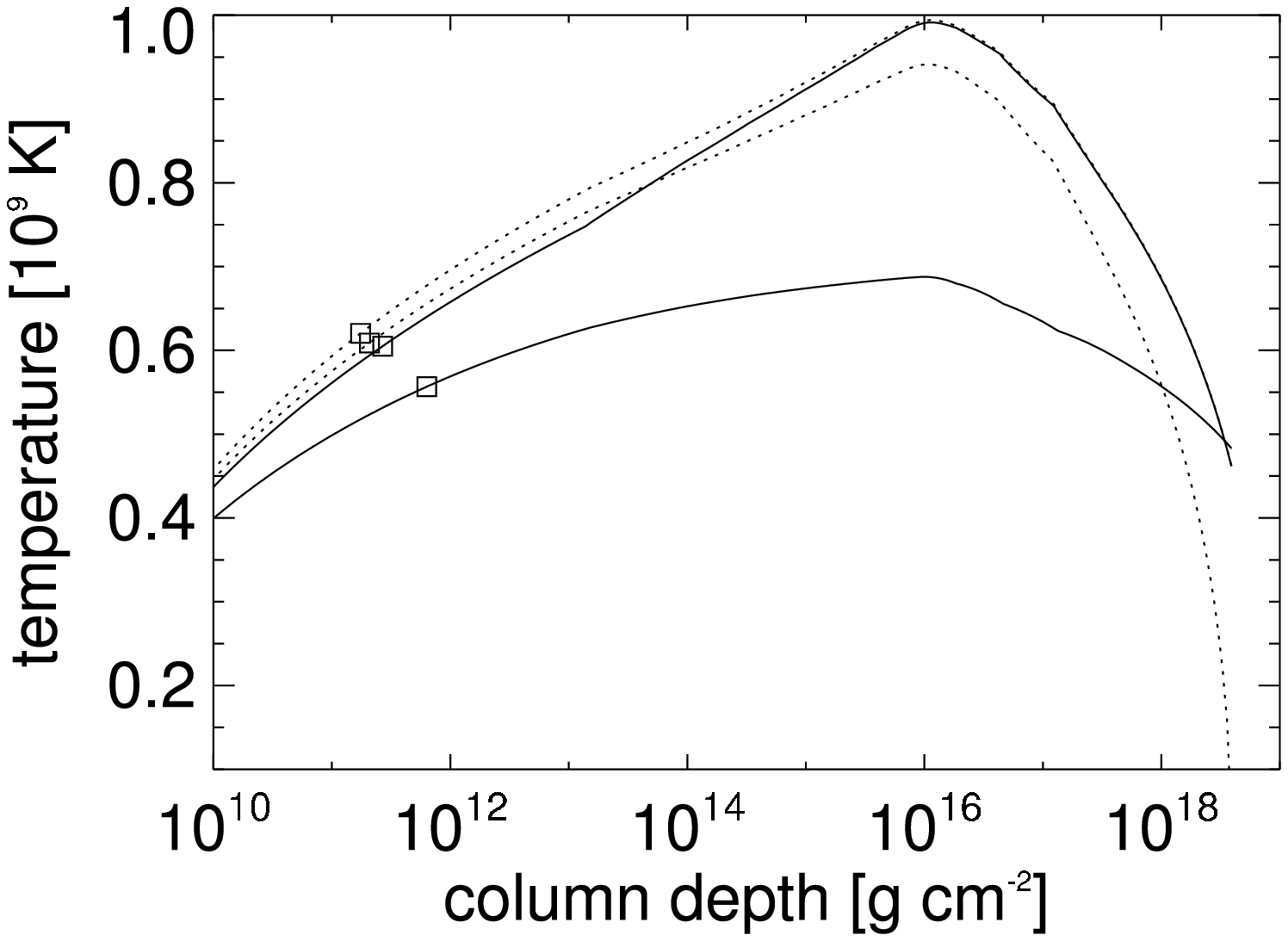}
  \caption{Same as \protect\ref{fig:mdot0.3}, but for $\Mdot =
  10^{18}\nsp\grampersecond$.}
  \label{fig:mdot1.0}
\end{figure}

The thermal profiles for these four cases are shown in
Fig.~\ref{fig:mdot0.3} for $\dot{M} = 3\ee{17}\nsp\gram\usp\second^{-1}$
(approximately \onethird\ Eddington) and Fig.~\ref{fig:mdot1.0} for
$\dot{M} = 10^{18}\nsp\gram\usp\second^{-1}$ (approximately Eddington).
The solutions with the higher core temperatures are for modified Urca
processes; the ones with lower core temperatures for direct Urca
processes.  For each core temperature there are two profiles, one for a
disordered, amorphous crust (\emph{solid lines}) and one for an impure
lattice (\emph{dotted lines}).

The squares mark the point at which we estimate unstable ignition to
occur.  To define ignition, we use the criteria from
\citet{cumming.bildsten:carbon} and set $\varepsilon_{1212} = \xi \rho K
T/y^2$, where \ECC\ is the mass-specific heating rate from the reaction
\CC, $\Ecool = \rho K T/y^2$ is an estimate of the cooling rate, and
$\xi \approx 2/26$ is the ratio
$(\partial_T\ln\Ecool)/(\partial_T\ln\ECC)$.  This simple approach gives
us a rough estimate of where ignition occurs and is known for X-ray
bursts to give results that are roughly in line with numerical
simulations \citep{woosley.heger.ea:models} and linear stability
analyses \citep{narayan.heyl:thermonuclear}.  For the \CC\ rate, we use
the formula of \citet{caughlan88:_therm} and incorporate strong
screening according to \citet*{ogata93:_therm}.

For $\Mdot = 3.0\ee{17}\nsp\grampersecond$, the recurrence times for
\CC\ ignition are, from the hottest profile to the coldest, 1.4, 3.0,
5.1, and 28\nsp\yr\ with total burst energies of $7.5\ee{41}$,
$1.6\ee{42}$, $2.7\ee{42}$ and $1.5\ee{43}\nsp\ergspersecond$,
respectively.  Likewise, for $\Mdot = 10^{18}\nsp\grampersecond$, the
recurrence times are 0.08, 0.10, 0.30, and 1.4\nsp\yr\ and the total
burst energies are $1.5\ee{41}$, $1.7\ee{41}$, $5.3\ee{41}$, and
$2.5\ee{42}\nsp\ergspersecond$.  \emph{Neutron stars with enhanced
  neutrino cooling are not compatible with observed superburst
  recurrence times and energetics unless the crust is amorphous.}  Note
that if the inner crust were to form a lattice with a low impurity
concentration, $Q\sim 1$, then the \CC\ ignition approaches the case
discussed by \citet{taam78:_nuclear}, with recurrence times of
10--100\nsp\yr.  For the case of a disordered lattice (\emph{solid
  curves}), the ignition conditions become nearly independent of the
core temperature at higher accretion rates.

\section{Conclusions}\label{sec:conclusions}

Using parameterized models of the NS core, we have computed ignition
conditions for \nuclei{12}{C} that solve for the flux emergent from the
inner crust.  There are two principal conclusions.  First, the flux in
the carbon-enriched layer scales roughly as $\meanA/\meanZZ$, so that
the superburst ignition is only weakly dependent on the composition of
that layer.  Second, we confirm the scenario of
\citet{cumming.bildsten:carbon}, but only if the crust thermal
conductivity is low.  In particular, the calculated recurrence times are
too long if there is \emph{both} enhanced core neutrino emission
\emph{and} the crust thermal conductivity is set by phonon and impurity
scattering.  This is relevant not only to studies of neutron star
cooling, but also to studies of the magnetic field evolution of the
crust \citep*[for a recent study, see][]{cumming.arras.ea:magnetic}.

In addition to the recurrence time and burst energetics, the cooling of
the NS atmosphere, as evidenced by the quenching of subsequent type I
X-ray bursts, can be used to infer $y_\mathrm{ign}$, the column depth of
the superburst ignition \citep{cumming.macbeth:thermal}.  The ignition
curve, as estimated in \S~\ref{sec:results}, gives a unique temperature
$T_\mathrm{ign}(y_\mathrm{ign})$.  Using this information thus reduces
the uncertainty from the poorly constrained mass accretion rate.  From
Fig.~\ref{fig:mdot0.3}, $y_\mathrm{ign}$ differs by a factor of 3--4
between different crust models for a core with only modified Urca
processes.

Perhaps the most intriguing superburst source is \source{KS}{1731}{-260}
\citep{kuulkers.ea:ks1731-superburst}, for which there are independent
estimates of the interior temperature and crust thermal conductivity.
This source was observed to be persistent for $\approx 12\nsp\yr$ until
it suddenly went into quiescence \citep{wijnands:ks1731}.
\citet{rutledge.ea.01:ks1731} noted that because of its long outburst,
the crust would be driven out of thermal equilibrium with the core, and
therefore follow-up observations of the quiescent X-ray luminosity would
measure the thermal relaxation of the crust.  Such observations
\citep{wijnands.ea:xmm_1731} revealed a rapid and strong decline of the
X-ray luminosity.  If the quiescent emission is solely powered by the
cooling crust, this sharp decrease in the quiescent luminosity suggests
that the crust has a high thermal conductivity, consistent with being a
locally pure lattice, and that the neutrino emission from the core is
stronger than modified Urca.  The calculations in this paper, being
steady-state, are not directly applicable to \source{KS}{1731}{-260},
but do indicate that the low quiescent emission, combined with the
occurrence of a superburst, are a challenge to our understanding of
accreting neutron stars.

\acknowledgements

We thank Andrew Cumming, Bob Rutledge, Hendrik Schatz, and Alexander
Heger for helpful conversations and comments.

\end{document}